\documentclass[prd,twocolumn,floatfix,preprintnumbers,letterpaper]{revtex4}
\usepackage{amssymb,amsbsy,bm,amsmath,psfrag,dsfont}
\usepackage{hyperref}
\usepackage{graphicx}
\usepackage{epsfig}

\begin{document}

\newcommand{\be}{\begin{equation}}
\newcommand{\ee}{\end{equation}}
\newcommand{\bea}{\begin{eqnarray}}
\newcommand{\eea}{\end{eqnarray}}
\newcommand{\no}{\noindent}
\newcommand{\la}{\lambda}
\newcommand{\si}{\sigma}
\newcommand{\vp}{\mathbf{p}}
\newcommand{\vk}{\vec{k}}
\newcommand{\vx}{\vec{x}}
\newcommand{\om}{\omega}
\newcommand{\Om}{\Omega}
\newcommand{\ga}{\gamma}
\newcommand{\Ga}{\Gamma}
\newcommand{\gaa}{\Gamma_a}
\newcommand{\al}{\alpha}
\newcommand{\ep}{\epsilon}
\newcommand{\app}{\approx}
\newcommand{\nc}{\newcommand}

\title{Astrophysical Constraints on Dark Energy}
\author{Chiu Man Ho} \email{cmho@msu.edu}
\affiliation{Department of
Physics and Astronomy, Michigan State University, East Lansing, MI 48824, USA}
\author{Stephen~D.~H.~Hsu} \email{hsu@msu.edu}
\affiliation{Department of
Physics and Astronomy, Michigan State University, East Lansing, MI 48824, USA}
\date{\today}

\begin{abstract}
Dark energy (i.e., a cosmological constant) leads, in the Newtonian approximation, to a repulsive force which grows linearly with distance and which can have astrophysical consequences. For example, the dark energy force overcomes the gravitational attraction from an isolated object (e.g., dwarf galaxy) of mass $10^7 M_\odot$ at a distance of $~ 23$ kpc. Observable velocities of bound satellites (rotation curves) could be significantly affected, and therefore used to measure or constrain the dark energy density. Here, {\it isolated} means that the gravitational effect of large nearby galaxies (specifically, of their dark matter halos) is negligible; examples of isolated dwarf galaxies include Antlia or DDO 190.
\end{abstract}
\maketitle

\section{Introduction}

The discovery of dark energy, which accounts for the majority of the energy in the universe, is one of the most significant of the last 20 years. While the repulsive properties of dark energy are well known in the cosmological context, they have not been as thoroughly understood on shorter, astrophysical, length scales. Previous work has constrained the cosmological constant on solar-system scales \cite{solar}, but its effects are obviously too small to be directly observed.

In what follows, we discuss the repulsive dark energy force and its astrophysical effects on galactic scales. Because this force grows linearly with distance, its effect is most likely to be significant for weakly-bound satellites with large orbits. To detect the effects of dark energy, we must first understand orbits resulting from ordinary gravitational dynamics, with potentials mainly determined by the distribution of dark matter. We find that observations of distant satellites of isolated dwarf galaxies could be used to detect the effects of dark energy. Here, isolated means sufficiently far from other sources of gravitational potential. When dwarf galaxy systems are not sufficiently isolated, the orbits of their satellites are subject to tidal forces from nearby large galaxies. These tidal forces can distort orbital shapes, and enforce an upper limit on orbital radii.

\section{Newtonian gravity and cosmological constant}

The Einstein equation with cosmological constant $\Lambda$ is
\bea
R_{\mu\nu} -\frac12\, g_{\mu\nu}\, R = 8\,\pi\, G\, T_{\mu\nu}+g_{\mu\nu}\,\Lambda\,.
\eea
Contracting both sides with $g^{\mu\nu}$, one gets $R = - 8\,\pi\, G\, T - 4 \, \Lambda $
where $T \equiv T^{\mu}_{~\mu}$ is the trace of the matter (including dark matter) energy-momentum tensor. This can be substituted in the original
equation to obtain
\bea
R_{\mu\nu} = 8\,\pi\, G\, \left(\,T_{\mu\nu}- \frac12\, g_{\mu\nu}\,T\,\right) - g_{\mu\nu}\,\Lambda\,.
\eea

In the Newtonian limit, one can decompose the metric tensor as $g_{\mu\nu}=\eta_{\mu\nu}+h_{\mu\nu}$ with $|h_{\mu\nu}| \ll 1$. Specifically, we are interested in the $00$-component of the Einstein equation. We parameterize the
$00^{\textrm{th}}$-component of the metric tensor as
\bea
\label{g00}
g_{00} = 1 + 2\, \Phi\,,
\eea
where $\Phi$ is the Newtonian gravitational potential. To leading order, one can show that \cite{Weinberg}
\bea
R_{00} \approx \frac12\,\vec{\nabla}^2\, g_{00} = \vec{\nabla}^2\, \Phi\,.
\eea
In the inertial frame of a perfect fluid, its 4-velocity is given by  $u_\mu = (1, \vec{0})$ and we have
\bea
T_{\mu \nu} = (\,\rho+ p\,) \,u_\mu \,u_\nu - p \,g_{\mu \nu} = {\rm diag} (\, \rho \, , \, p \,)\,,
\eea
where $\rho$ is the energy density and $p$ is the pressure. For a Newtonian (non-relativistic) fluid, the pressure is negligible compared
to the energy density, and hence $T \approx T_{00} =\rho$. As a result, in the Newtonian limit, the $00$-component of the Einstein equation reduces to
\bea
\label{Poisson}
\vec{\nabla}^2\, \Phi = 4\,\pi\, G\,\rho - \Lambda\,,
\eea
which is just the modified Poisson equation for  Newtonian gravity, including cosmological constant. This equation can also be derived from the Poisson equation of Newtonian gravity,  $\vec{\nabla}^2\, \Phi = 4\,\pi\, G\, ( \rho + 3p )$, with source terms from matter and dark energy; $p \approx 0$ for non-relativistic matter, and $p = -\rho$ for a cosmological constant.

Assuming spherical symmetry, we have $\vec{\nabla}^2\, \Phi = \frac{1}{r^2}\,\frac{\partial}{\partial r}\left(\,r^2\,\frac{\partial \Phi}{\partial r}\,\right)$ and the Poisson equation is easily solved to obtain
\bea
\label{potential}
\Phi = -\frac{G M}{r} -\frac{\Lambda}{6}\,r^2\,,
\eea
where $M$ is the total mass enclosed by the volume $\frac43 \pi r^3$. The corresponding gravitational field strength
is given by
\bea
\label{fieldstrength}
\vec{g} = -\vec{\nabla}\,\Phi = \left(\,-\frac{G M}{r^2} +\frac{\Lambda}{3}\,r\,\right)\,\hat{r}\,.
\eea
Therefore, the cosmological constant leads to a repulsive force whose strength grows linearly with $r$.

One can also derive $\vec{g}$ by starting with the de Sitter-Schwarzschild metric \cite{dSS}
\bea
ds^2 &=& \left(\,1- \frac{2 G M}{r}-\frac{\Lambda}{3}r^2\,\right)\,dt^2  \\
&& ~~ - \left(\,1- \frac{2 G M}{r}-\frac{\Lambda}{3}r^2\,\right)^{-1}\,dr^2  ~-~ r^2 d\Omega^2 \nonumber \\
\nonumber
\eea
which describes the spacetime outside a spherically symmetric mass distribution $M$ in the presence of a cosmological constant
$\Lambda$. One then obtains Eq. \eqref{potential} and hence Eq. \eqref{fieldstrength} by identifying Eq. \eqref{g00} with the $00$-component of the de Sitter-Schwarzschild metric.

\section{Galaxies}

The results obtained in the previous section are relevant to galaxies. For instance, in the presence of the cosmological constant $\Lambda$,
Eq. \eqref{fieldstrength} describes the Newtonian gravitational field strength outside a galaxy with a spherically symmetric mass
distribution $M$. From Eq. \eqref{fieldstrength}, it is clear that when $r$ is sufficiently large, the repulsive dark force will dominate over the gravitational attraction. The critical value of r beyond which this happens is given by
\bea
\label{critical}
r_c = \left(\,\frac{3 \, G\,M} {\Lambda}\,\right)^{1/3} = \left(\,\frac{3 \,M} {8\,\pi\,\rho_{\Lambda}}\,\right)^{1/3}\,,
\eea
where $\rho_{\Lambda} = \frac{\Lambda}{8\,\pi\,G} \approx (\,2.3\times 10^{-3} \textrm{eV}\,)^4$ is the observed energy density of the cosmological
constant. Table I displays galactic masses in units of solar mass $M_\odot$ and the corresponding $r_c$.

\begin{table}[t!]
  \centering
  \begin{tabular}{l*{4}{c}r}
Galaxy Mass              ~~~~~  &  $r_c$  \\ \hline \hline
$10^6 M_\odot$           ~~~~~  &    10.7 kpc         \\ \hline
$10^7 M_\odot$           ~~~~~  &    23.1 kpc         \\ \hline
$10^8 M_\odot$           ~~~~~  &    49.8 kpc         \\ \hline
$10^9 M_\odot$           ~~~~~  &    107 kpc         \\ \hline
$10^{10} M_\odot$        ~~~~~  &    231 kpc        \\ \hline
$10^{11} M_\odot$        ~~~~~  &    498 kpc        \\ \hline
$10^{12} M_\odot$        ~~~~~  &    1.07 Mpc         \\ \hline
$10^{13} M_\odot$        ~~~~~  &    2.31 Mpc         \\ \hline
$10^{14} M_\odot$        ~~~~~  &    4.98 Mpc         \\ \hline
\end{tabular}
 \caption{~Galaxy masses (units of solar mass $M_\odot$) and the corresponding $r_c$.}
\label{table1}
\end{table}

Typical galaxies, including our Milky Way, have total mass (including dark matter) $\gtrsim 10^{11-12} \, M_\odot$ and sizes $\sim \, 50\, \textrm{kpc}$. According to Table I, $r_c \gtrsim 500 \,\textrm{kpc}$ for these galaxies, so the dark force is not likely to affect internal dynamics, but may impact galaxy-galaxy interactions \cite{galaxy-galaxy}, and limit the size of galaxy clusters ($\sim \, 10^{14} M_\odot$,
size $\sim \,$ Mpc).

Some dwarf galaxies have total mass (including dark matter) $\sim 10^{7} M_\odot$. These include Ursa Major II, Coma Berenices,
Leo T, Leo IV, Canes Venatici I, Canes Venatici II, and Hercules (analyzed by \cite{SimonGeha}), and also Leo II \cite{LeoII} and
Leo V \cite{LeoV}. The irregular galaxies Leo A \cite{LeoA}, Antlia \cite{Antlia} and DDO 190 \cite{DDO190} also have masses around
$10^{7} M_\odot$.

For galaxies
with mass $\sim 10^{7} M_\odot$, we have $r_c \sim  23 \,\textrm{kpc}$. Thus, their galactic rotation curves could be affected by the dark force: rotational velocities of stars or gas clouds bound to
these galaxies should be smaller than that predicted by ordinary Newtonian gravity. This in turn could provide
a novel way to measure the cosmological constant in the future. Rotation curves for many galaxies have been measured to radii of $\sim 30 \,\textrm{kpc}$ or more, and for some dwarf galaxies to $\sim 10 \,\textrm{kpc}$ \cite{swaters}. Low surface brightness (LSB) galaxies may also be worthy of investigation \cite{LSB}. Some LSBs with total mass $\sim \, 10^{10} M_\odot$ have disks as large as $100 \,\textrm{kpc}$.

The Navarro-Frenk-White (NFW) profile \cite{NFW} is a commonly used parametrization of dark matter halo energy density:
\bea
\label{NFW}
\rho = \frac{\rho_0}{ \, r /R_s \,\left(1+  r / R_s  \right)^2}\,,
\eea
where $\rho_0$ is a characteristic halo density and $R_s$ is the scale radius. These two quantities vary from galaxy to galaxy. While the detailed shape of the actual dark matter density may differ from the NFW profile, the asymptotic $1/r^3$ behavior is widely accepted. Our results below will not be sensitive to the density profile at small $r$.

Consider a dwarf galaxy (DG) and a larger galaxy (LG) (e.g., the Milky Way) whose centers of mass are separated by a distance $R$, and a satellite of the DG whose orbital radius is roughly $r$. If the distance $R$ is sufficiently large, we can neglect the gravitational potential of the LG and treat the DG-satellite system as approximately isolated. In that case, the values in Table I provide a rough guide for distances $r$ at which the dark energy force becomes significant. In the following section we will investigate to what extent measurement of satellite velocities can constrain the dark energy density around the DG.

But first let us examine in more detail under what circumstances we can neglect the gravitational effects from the (dark matter halos) of
neighboring galaxies on the DG. We will assume an NFW profile for both the DG halo and the larger galactic halo. The distance $R$ from the
center of the DG to the center of the LG is generally not equal to the distance from the satellite to the center of the LG, which can vary
from $(R-r)$ to $(R+r)$. Therefore, the gravitational pull exerted on the satellite by the LG is different from the pull on the DG, leading
to a tidal effect. (See \cite{Tidal} for previous work regarding the tidal effects on orbiting satellites around their host galaxies.) This tidal
effect is repulsive: it pulls apart the DG-satellite system. Perhaps surprisingly, for many DGs (i.e., near the Milky Way), the tidal effect is
large enough to distort and even destabilize the satellite orbits.

\begin{table}[t!]
  \centering
  \begin{tabular}{l*{4}{c}r}
Dwarf Galaxies           ~~  &  distance to MW   ~~ &   distance to M31    \\ \hline \hline
Leo T                    ~~  &    422 kpc        ~~ &        991 kpc      \\ \hline
Leo IV                   ~~  &    155 kpc        ~~ &        899 kpc      \\ \hline
Canes Venatici I         ~~  &    218 kpc        ~~ &        864 kpc      \\ \hline
Canes Venatici II        ~~  &    161 kpc        ~~ &        837 kpc      \\ \hline
Hercules                 ~~  &    126 kpc        ~~ &        826 kpc      \\ \hline
Leo II                   ~~  &    236 kpc        ~~ &        901 kpc      \\ \hline
Leo V                    ~~  &    179 kpc        ~~ &        915 kpc      \\ \hline
Leo A                    ~~  &    803 kpc        ~~ &        1200 kpc      \\ \hline
Antlia                   ~~  &    1350 kpc        ~~ &       2039 kpc      \\ \hline
DDO 190                  ~~  &    2793 kpc        ~~ &       2917 kpc      \\ \hline
\end{tabular}
\caption{~Some dwarf galaxies with mass $\sim 10^{7} M_\odot$ and their distances
from the Milky Way (MW) and the Andromeda galaxy (M31). See Table 2 in \cite{DGreview}. }
\label{table2}
\end{table}

\begin{table}[t!]
  \centering
  \begin{tabular}{l*{4}{c}r}
Dwarf Galaxies           ~~  &  $r_{\textrm{half}}$   ~~ &   $\rho_0/\,\textrm{GeV cm}^{-3}$   \\ \hline \hline
Leo T                    ~~  &    178 $\pm$ 39 pc        ~~ &  \{0.028,\,0.22,\,1.74\}   \\ \hline
Leo IV                   ~~  &    116 $\pm$ 30 pc        ~~ &  \{0.037,\,0.25,\,1.66\}           \\ \hline
Canes Venatici I         ~~  &    564 $\pm$ 36 pc        ~~ &  \{0.017,\,0.21,\,2.69\}              \\ \hline
Canes Venatici II        ~~  &    74 $\pm$ 12 pc        ~~ &   \{0.053,\,0.30,\,1.65\}          \\ \hline
Hercules                 ~~  &    330 $\pm$ 63 pc        ~~ &  \{0.020,\,0.20,\,2.05\}          \\ \hline
Leo II                   ~~  &    123 $\pm$ 27 pc        ~~ &  \{0.035,\,0.24,\,1.67\}         \\ \hline
Leo V                    ~~  &    42 $\pm$ 5 pc          ~~ &  \{0.086,\,0.38,\,1.72\}           \\ \hline
Leo A                    ~~  &    354 $\pm$ 19 pc        ~~ &  \{0.019,\,0.20,\,2.11\}          \\ \hline
Antlia                   ~~  &    471 $\pm$ 52 pc      ~~ &    \{0.018,\,0.21,\,2.42\}         \\ \hline
DDO 190                  ~~  &    520 $\pm$ 49 pc       ~~ &   \{0.017,\,0.21,\,2.56\}          \\ \hline
\end{tabular}
\caption{~Some dwarf galaxies with mass $\sim 10^{7} M_\odot$ and their half-light radii $r_{\textrm{half}}$
and $\rho_0$. For each galaxy, the three different values of $\rho_0$ correspond to the exponents
$\{-1.6-0.4,\, -1.6,\, -1.6+0.4\}$ in Eq. \eqref{meanrho}. }
\label{table3}
\end{table}

\begin{table}[t!!]
  \centering
  \begin{tabular}{l*{4}{c}r}
Dwarf Galaxies           ~~  &  negligible tidal effect       \\ \hline \hline
Leo T                    ~~  &   $ r \,\lesssim\, $ $5.6^{+5.5}_{-2.8}$ kpc              \\ \hline
Leo IV                   ~~  &    $ r \,\lesssim\, $ $1.9^{+1.7}_{-0.9}$ kpc             \\ \hline
Canes Venatici I         ~~  &   $ r \,\lesssim\, $ $2.6^{+3.5}_{-1.5}$ kpc          \\ \hline
Canes Venatici II        ~~  &    $ r \,\lesssim\, $ $2.2^{+1.7}_{-0.9}$ kpc            \\ \hline
Hercules                 ~~  &    $ r \,\lesssim\, $ $1.4^{+1.6}_{-0.8}$ kpc           \\ \hline
Leo II                   ~~  &   $ r \,\lesssim\, $ $3.1^{+2.8}_{-1.5}$ kpc             \\ \hline
Leo V                    ~~  &    $ r \,\lesssim \,$ $2.8^{+1.8}_{-1.1}$ kpc            \\ \hline
Leo A                   ~~  &   ~ $ r \,\lesssim \,$ $10.6^{+12.6}_{-5.8}$ kpc            \\ \hline
Antlia                   ~~  &   ~ $ r \,\lesssim \,$ $18.8^{+23.8}_{-10.6}$ kpc            \\ \hline
DDO 190                  ~~  &   ~~ $ r \,\lesssim \,$ $39.4^{+51.3}_{-22.4}$ kpc              \\ \hline
\end{tabular}
\caption{~ Required orbital radii for the satellites of some dwarf galaxies with mass $\sim 10^{7} M_\odot$ to ensure negligible tidal effect.
The upper limit on $r$ is determined by requiring $F_{\textrm{LG}}^{\textrm{tidal}}/F_{\textrm{DG}} \lesssim 0.1$. The central value corresponds to the exponent $-1.6$ in Eq. \eqref{meanrho} while the $\pm$ values correspond to the exponents $-1.6\pm 0.4$.}
\label{table4}
\end{table}

Let the total mass of the DG enclosed within $r$ be $M_{\textrm{DG}}(r)$ and that of the LG enclosed within $R\pm r$ be $M_{\textrm{LG}}(R\pm r)$.
Then we have
\bea
\label{MDG}
M_{\textrm{DG}}(r) &=& \int_{0}^{r}\, 4\,\pi\, {r'}^2\, \rho_{\textrm{DG}}(r') \, dr' \,,\\
\label{MLG}
M_{\textrm{LG}}(R\pm r) &=& \int_{0}^{R\pm r}\, 4\,\pi\, {r'}^2\, \rho_{\textrm{LG}}(r') \, dr' \,.
\eea
The circularity and stability of the satellite orbits can be guaranteed by requiring that the magnitude of the tidal force due to the LG, $F_{\textrm{LG}}^{\textrm{tidal}}$, is much smaller than the gravitational pull due to DG, $F_{\textrm{DG}}$. This requires
\bea
F_{\textrm{LG}}^{\textrm{tidal}}  \,\approx\, \frac{G\,M_{\textrm{LG}}(R)}{R^2}\, \frac{r}{R} ~ \ll ~ F_{\textrm{DG}} \,=\, \frac{G\,M_{\textrm{DG}}(r)}{r^2}\,,
\eea
which implies
\bea
\label{condition}
\left(\,\frac{M_{\textrm{LG}}(R)}{M_{\textrm{DG}}(r)}\,\right)^{1/3}\, \frac{r}{R} ~\ll~ 1\,.  \\ \nonumber
\eea

According to \cite{DGreview}, many dwarf galaxies are at least 100 kpc away from the Milky Way and much farther from
the Andromeda galaxy (M31). Some of these dwarf galaxies with mass $\sim 10^{7} M_\odot$ include Leo T, Leo IV, Canes Venatici I,
Canes Venatici II, Hercules, Leo II, Leo V, Leo A, Antlia and DDO 190. Their distances from the Milky Way and M31 are shown in Table II.

For the Milky Way (MW), we have $\rho_0 \sim 0.2$ GeV cm$^{-3}$ and $R_s \sim 25$ kpc (see Fig. 1 in \cite{Poor}).
For dwarf galaxies with mass $\sim 10^{7} M_\odot$, the analysis in \cite{Walker} suggests that $R_s \sim 0.795 \,\textrm{kpc}$
(see their Table 3 which gives the best fit parameters for some dwarf galaxies assuming the NFW profile).
In terms of the half-light radius $r_{\textrm{half}}$ (the radius at which half of the total light is emitted), \cite{Walker} obtains
a relation for the mean density $\langle \rho \rangle$ interior to $r_{\textrm{half}}$:
\bea
\label{meanrho}
\langle \rho \rangle \,\sim\, 2600\, \left(\,\frac{r_{\textrm{half}}}{\textrm{pc}}\,\right)^{-1.6 \pm 0.4}\,\,\,\textrm{GeV cm}^{-3}\,.
\eea
(Note that $M_\odot$ pc$^{-3}$ $\approx$ 40 GeV cm$^{-3}$). In the second column of Table III, the half-light radii of some dwarf galaxies with mass $\sim 10^{7} M_\odot$ are listed. We adopt these values of $r_{\textrm{half}}$ from \cite{DGreview,Walker,Bullock}. For each of the galaxies listed in Table III, by using $r = r_{\textrm{half}}$ and the corresponding value of $\langle \rho \rangle$ obtained from Eq. \eqref{meanrho} (together with $R_s \sim 0.795 \,\textrm{kpc}$) in the NFW density profile, we estimate the value of $\rho_0$. The three different values of $\rho_0$ listed in the third column of Table III correspond to the exponents $\{-1.6-0.4,\, -1.6,\, -1.6+0.4\}$ in Eq. \eqref{meanrho}.

Using the normalization factors ($\rho_0$ and $R_s$) for both the Milky Way and the dwarf galaxies with mass $\sim 10^{7} M_\odot$, we can determine the satellite radii for which Eq. \eqref{condition} is satisfied. In Table IV, we display the required orbital radii for the satellites of some dwarf galaxies with mass $\sim 10^{7} M_\odot$, assuming negligible tidal effect. The upper limit on $r$ is determined by requiring $F_{\textrm{LG}}^{\textrm{tidal}}/F_{\textrm{DG}} \lesssim 0.1$. The central value corresponds to the exponent $-1.6$ in
Eq. \eqref{meanrho} while the $\pm$ values correspond to the exponents $-1.6\pm 0.4$.

\section{Constraints on dark energy density from rotation curves}

For galaxies with mass $\sim 10^{7} M_\odot$, we have $r_c \sim  23 \,\textrm{kpc}$, so the effect of the dark energy force becomes observationally significant for $r \sim 20 \,\textrm{kpc}$. For such radii, Table IV indicates that the tidal effect due to the Milky Way might have already distorted and even destabilized the satellite orbits in the dwarf galaxies Leo T, Leo IV, Canes Venatici I, Canes Venatici II, Hercules, Leo II, Leo V and Leo A. On the other hand, satellite orbits of Antlia and DDO 190 could be strongly affected by the dark energy force at radii for which the tidal effect due to the Milky Way is negligible.

\begin{table}[t!]
  \centering
  \begin{tabular}{l*{4}{c}r}
$N$     ~~  &  orbital radii   ~~ &  95\% CI of\,\, $c / 10^{-84}\,\textrm{GeV}^2$   \\ \hline \hline
5       ~~  &    1-10 kpc        ~~ &       \{\,1.31,\, 1.78\,\}        \\ \hline
5       ~~  &    5-15 kpc        ~~ &        \{\,1.49,\, 1.66\,\}    \\ \hline
5       ~~  &    10-20 kpc        ~~ &        \{\,1.54,\, 1.61\,\}     \\ \hline
5       ~~  &    15-25 kpc        ~~ &        \{\,1.55,\, 1.60\,\}     \\ \hline
5       ~~  &    20-30 kpc        ~~ &        \{\,1.56,\, 1.59\,\}    \\ \hline
10      ~~  &    1-10 kpc        ~~ &      \{\,1.43,\, 1.70\,\}       \\ \hline
10      ~~  &    5-15 kpc        ~~ &        \{\,1.53,\, 1.64\,\}    \\ \hline
10      ~~  &    10-20 kpc        ~~ &        \{\,1.56,\, 1.61\,\}      \\ \hline
10      ~~  &    15-25 kpc        ~~ &        \{\,1.56,\, 1.60\,\}    \\ \hline
10      ~~  &    20-30 kpc        ~~ &        \{\,1.57,\, 1.59\,\}     \\ \hline
\end{tabular}
\caption{~ 95\% confidence interval (CI) of $c$, assuming 1\% error in $v^2$. $N$ is number of satellites.}
\label{table5}
\end{table}

\begin{table}[t!]
  \centering
  \begin{tabular}{l*{4}{c}r}
$N$     ~~  &  orbital radii   ~~ &  95\% CI of\,\, $c / 10^{-84}\,\textrm{GeV}^2$   \\ \hline \hline
5       ~~  &    1-10 kpc        ~~ &        \{\,0.40,\,  2.31\,\}    \\ \hline
5       ~~  &    5-15 kpc        ~~ &        \{\,1.11,\, 1.81\,\}     \\ \hline
10      ~~  &    1-10 kpc        ~~ &      \{\,0.68,\, 2.07\,\}       \\ \hline
10      ~~  &    5-15 kpc        ~~ &        \{\,1.25,\, 1.78\,\}    \\ \hline
\end{tabular}
\caption{~ 95\% confidence interval (CI) of $c$, assuming 5\% error in $v^2$. $N$ is number of satellites.}
\label{table6}
\end{table}

For isolated dwarf galaxies such as Antlia and DDO 190, the rotational velocity-squared $v^2$ of their satellites at $r$ is given by
\bea
v^2(r) = \frac{G\, M_{\textrm{DG}}(r)}{r} -\frac{1}{3}\,\Lambda\,r^2\,,
\eea
where $M_{\textrm{DG}}(r)$ can be obtained by a simple integration:
\bea
M_{\textrm{DG}}(r) = 4\,\pi\,\rho_0\,R_s^3\,\left[\,\ln\left(\,\frac{r+R_s}{R_s}\,\right)-\frac{r}{r+R_s}\,\right]\,.
\eea
Thus, for a set of measurements on $v^2(r)$ at some level of sensitivity, we fit $v^2(r)$ with
\bea
v^2(r) = \frac{a}{r} \,\left[\,\ln\left(\,\frac{r+b}{b}\,\right)-\frac{r}{r+b}\,\right] - c\,r^2\,,
\eea
where $a$ and $b$ are some constants. A (statistically significant) positive fit value for $c$ suggests the existence of a cosmological constant. The cosmological value for $c$ is $c=1.58 \times 10^{-84}$ GeV$^2$. (Note that in a realistic situation we cannot rely on the NFW parametrization being exact, and the fitting function should probably be slightly more general than the one used above.)

In Table V, we simulate the results of measurements on $v^2(r)$ with corresponding error of 1\%. We take $\rho_0 \sim 0.2\, \textrm{GeV cm}^{-3}$
and $R_s \sim 0.795 \,\textrm{kpc}$ for the dwarf galaxies. We vary the number of satellites $N$
and their (randomly generated) orbital radii. For example, at 95\% confidence level, one could bound $c$ to be positive using 5 satellites at
$r \sim 1-10$ kpc. In order to bound $c$ close to its cosmological value, one would need, e.g.,
at least 5 satellites at $r \sim 10-20$ kpc or 10 satellites at $r \sim 5-15$ kpc.

In the event that dark energy is dynamical \cite{dynamical}, as opposed to a rigid cosmological constant, it might form inhomogeneous clumps on
galactic length scales. This behavior could, in principle, be detectable through the effects discussed here: $\Lambda$ from astrophysical
measurements would be larger than the known cosmological value. In Table VI, we simulate the results from measurements on $v^2(r)$, assuming
that the corresponding error is 5\%. Again, we take $\rho_0 \sim 0.2\, \textrm{GeV cm}^{-3}$ and $R_s \sim 0.795 \,\textrm{kpc}$ for the dwarf
galaxies. The table indicates that even at the sensitivity of 5\%, one could rule out (at 95\% confidence level) any $\Lambda$ that is
significantly larger than $1.58 \times 10^{-84}$ GeV$^2$ by using, e.g., 5 satellites at $r \sim 1-10$ kpc. The very existence of satellites of dwarf galaxies (even those close to the Milky Way, and hence subject to significant tidal forces that limit $r$) provides an upper limit on the local dark energy density, probably no more than an order of magnitude larger than the cosmological value.

\section{Missing Satellite Problem}

Observations indicate fewer satellite galaxies than predicted by numerical simulations involving cold
dark matter. This is known as the missing satellite problem \cite{missing}. For instance, simulations predict a few hundred satellite galaxies
within a few $\textrm{Mpc}$ radius of the Local Group, but we have observed at least five times fewer.


For the Local Group with mass $\sim 10^{13-14} \, M_\odot$, we have $r_c \sim \,\textrm{a\, few\, Mpc}$, which naively suggests that the dark force might play a limiting role in the binding of satellite galaxies. However, if simulations of galaxy formation are performed with sufficient resolution and sufficiently late end-times in an expanding $\Lambda$CDM universe, the effect of the dark energy has already been accounted for (barring more exotic scenarios, like clumping of dark energy).

In other words, we are not suggesting the dark energy force discussed here as a possible solution to the missing satellite problem. We are instead suggesting a possible alternative method for measuring the local dark energy density through rotation curves of dwarf galaxies.

\bigskip

\emph{Acknowledgements.}\,\, We thank Robert Scherrer, James Schombert, Brian O'Shea, Megan Donahue, Jay Strader and Matthew Walker for useful
conversations. This work was supported by the Office of the Vice-President for Research and Graduate Studies at
Michigan State University.


\end{document}